\documentclass[useAMS,usenatbib,usegraphicx,times]{mn2e}

\def\ergps{erg~s$^{-1}$}
\def\kmpspMpc{km~s$^{-1}$~Mpc$^{-1}$}
\def\nH{$N_{\rm H}$\thinspace}
\def\psqcm{cm$^{-2}$}
\def\ergpspsqcm{erg~cm$^{-2}$~s$^{-1}$}
\def\cps{ct\thinspace s$^{-1}$}

\title[A Compton Thick AGN in IRAS 00182--7112]
 {A Compton Thick AGN Powering the Hyperluminous Infrared Galaxy IRAS 00182--7112}
\author[K. Nandra \& K. Iwasawa]
  {K. Nandra,$^1$ and K. Iwasawa$^2$ \\
$^1$Astrophysics Group, Imperial College London, Blackett Laboratory,
Prince Consort Road, London SW7 2AW \\
$^1$Max-Planck-Institut f\"ur extraterrestrische Physik, Giessenbachstrasse, 85748 Garching, Germany}

\date{2007 June 15}

\pagerange{\pageref{firstpage}--\pageref{lastpage}} \pubyear{2007}

\def\ergps{erg~s$^{-1}$}
\def\kmpspMpc{km~s$^{-1}$~Mpc$^{-1}$}
\def\nH{$N_{\rm H}$\thinspace}
\def\psqcm{cm$^{-2}$}
\def\ergpspsqcm{erg~cm$^{-2}$~s$^{-1}$}
\def\cps{ct\thinspace s$^{-1}$}

\usepackage{times}
\usepackage{graphicx}
\begin{document}

\label{firstpage}

\maketitle

\begin{abstract}
We present X-ray observations of the Hyperluminous Infrared Galaxy (HLIRG) IRAS 00182--7112 (F00183--7111) obtained using the XMM-Newton EPIC camera. A luminous hard X-ray source co-incident with the nucleus is revealed, along with weaker soft X-ray emission which may be extended or offset from the hard. The EPIC spectrum is extremely flat and shows Fe K$\alpha$ emission with very high equivalent width: both are typical characteristics of a buried, Compton--thick AGN which is seen only in scattered light. Perhaps the most remarkable characteristic of the spectrum is that the Fe K$\alpha$ line energy is that of He-like iron, making IRAS 00182--7112 the first hidden AGN known to be dominated by ionized, Compton thick reflection. Taking an appropriate bolometric correction we find that this AGN could easily dominate the FIR energetics. The nuclear reflection spectrum is seen through a relatively cold absorber with column density consistent with recent Spitzer observations. The soft X-ray emission, which may be thermal in nature and associated with star-forming activity, is seen unabsorbed.  The soft X-rays and weak PAH features both give estimates of the star formation rate $\sim 300 M_{\odot}$~yr$^{-1}$, insufficient to power the FIR emission and supportive of the idea that this HLIRG is AGN-dominated. 
\end{abstract}


\section{Introduction}

IRAS detected a new class of galaxy with large bolometric luminosity
dominated by the far-infrared (FIR), e.g., Sanders \& Mirabel
(1996). The most extreme objects in this class, indeed among the most
extreme in the universe, are the Hyper-luminous infrared
galaxies (HLIRGs: Rowan-Robinson 2000). With FIR luminosites in excess
of $10^ {13}L{\odot}$, their star formation rates would have to exceed 1000
$M_{\odot}$yr$^{-1}$ if powered solely by stellar processes. It is
known that a large fraction of luminous IR galaxies display emission
lines in their optical spectra characteristic of active galactic
nuclei (AGN), and this fraction rises as the bolometric
luminosity of the objects increases (Veilleux et al. 1995). There has been
considerable debate over whether the primary power source 
comes from the AGN (e.g., Sanders 1999) or star formation
(e.g., Genzel et al 1998). Given the abundant obscuration, it has
been extremely difficult to satisfactorily resolve this issue, and
indeed it is clear that both processes are usually important (e.g. Farrah et al. 2003).

New observations from space (ISO and Spitzer) and ground-based observatories, 
have brought about great progress in studying these very dusty objects, 
but the penetrating power of hard X-rays make them a powerful 
complementary probe. Accreting black holes are efficient X-ray emitters,
and the AGN in some luminous IR galaxies like NGC4945 and NGC6240 were found
first with hard X-ray observations (Iwasawa et al 1993; 
Done et al 1996; Iwasawa \& Comastri 1998; Vignati et al 1999; 
see Armus et al 2006 and Risaliti et al 2006 for evidence of active
nuclei based on infrared properties). These objects are very heavily
obscured and the absorption towards the X-ray nucleus is often Compton
thick, i.e. the column density \nH $>1/\sigma_{\rm T}\simeq 1.5\times 10^{24}$\psqcm, 
where $\sigma_{\rm T}$ is the Thomson cross section. If $N_{\rm H}$ greatly exceeds 
this value, even the direct hard X-rays may be blocked, but a hidden AGN can still be revealed through
its reflected light. In the typical bandpass ($<10$ keV) of high
sensitivity X-ray observatories like {\t Chandra} and
{\it XMM-Newton}, the spectrum is characterised by a flat continuum,
with photon index $\Gamma\sim0-1$ accompanied by a very strong
Fe K$\alpha $ emission line close to 6.4 keV. This characteristic spectrum has been observed in
many nearby Seyfert 2 galaxies (e.g., Matt et al 2000). At least
two classical HLIRGs, IRAS 09104+ 4109 (Kleinmann et al
1988) and IRAS F15307+3252 (Cutri et al 1994) have also been found to host
Compton thick AGN (Franceschini et al 2000; Iwasawa et al 2001;
Iwasawa et al 2005). Despite the difficulties in identifying Compton thick AGN they
are of widespread importance, as their precise numbers are crucial for
models of the X-ray background (e.g. Gilli et al. 2007), the accretion budget
of the universe (Fabian \& Iwasawa 1999) and last, but not least, the power source
in objects like HLIRGs. 

IRAS 00182--7112 (=IRAS F00183--7111) is an HLIRG discovered by IRAS
(Rowan-Robinson 2000 ) and optically identified at a redshift
$z=0.327$ (Saunders et al 1988). Unlike the two classical HLIRGs
mentioned above, which show Seyfert-2 characteristics in
their optical spectra, IRAS 00182--7112 has a colder IRAS colour and
has the optical spectrum of LINER, with no high-excitation emission-lines indicating the presence of AGN (Armus et al. 1989). Even the mid-infrared spectrum is devoid of
AGN emission lines (Spoon et al 2004). Instead it is
characterised by a deep slicate absorption and weak PAH features
(Spoon et al 2006). The continuum is carved by deep silicate,
hydrocarbon and various ice absorption features in the ISO and Spitzer
data (Spoon et al 2002, 2004), which suggests the power source, be it
an AGN or stars, must lie behind heavy obscuration.

The possible presence of an AGN is, however, suggested by a
mid-IR hot dust component similar to that in  NGC1068 
(e.g., Sturm et al 2005).  In addition, the compact radio core, 
which is not resolved at 1 arcsec resolution of the VLBI (Drake et al. 2004), 
has a flux significantly above the
FIR-radio correlation for star-forming galaxies (e.g., Appleton et al
2004), indicating the radio source might be powered by an AGN.
Here, we present XMM-Newton observation of this
HLIRG to investigate further the possible presence of an AGN and
the origin of the large infrared luminosity.

The currently popular cosmology with $H_0=73$ \kmpspMpc, $\Omega_{\rm
M}=0.27$, $\Omega_{\Lambda}=0.73$ is assumed in this paper. With
$z=0.327$, the luminosity and angular distances are $D_{\rm L}\simeq
1657$ Mpc and $D_{\rm A}\simeq 941$ Mpc and the angular scale is 4.5
kpc arcsec$^{-1}$. Because of the different cosmology assumed here
from the original definition of HLIRGs, the infrared luminosity of
IRAS 00182-- 7112 is $\sim 7\times 10^{12}L_{\odot}$, slightly under
$10^{13}L_{\odot}$.

\section{The XMM-Newton observation}

IRAS 00182--7112 was observed with XMM-Newton (ObsID 0147570101)  on 2003 April 17. The useful exposure times for the EPIC pn and MOS cameras are 10 ks and 12 ks, respectively, after discarding time intervals of high particle background in the 22 ks pointing duration. The data were taken in
full-frame mode with the thin filter in place. The data reduction was carried out using the stardard XMM SAS 7.0 and FTOOLS. The spectral analysis was performed using the XSPEC version 11.3 (Arnaud
et al 1996).

A source was detected at the position of the HLIRG with observed fluxes of $7\times 10^{-15}$\ergpspsqcm in the 0.5--2 keV band and $1.5\times 10^{-13}$\ergpspsqcm in the 2--10 keV band. The total 0.3-10 keV source counts obtained with the pn and two MOS cameras are 169 and
167 counts, respectively. Source spectra were extracted from a circle with radius 25$^{\prime}$, with background extracted from neighbouring source--free regions. The MOS1 and MOS2 spectra were co-added. We have adopted two different binnings for the spectra. When determining the emission line parameters below, to maintain sufficient resolution we rebinned the pn spectrum to a minimum of 12 counts per bin, and the MOS spectrum to 7 counts per bin, and determined the parameters using a maximum likelihood C-statistic (Cash 1976). When determining the broad-band properties we use a standard $\chi^{2}$ fit and have binned the spectra to a minimum of 20 counts per bin. 

\section{The X-ray images}

\begin{figure}
\includegraphics[width=84mm]{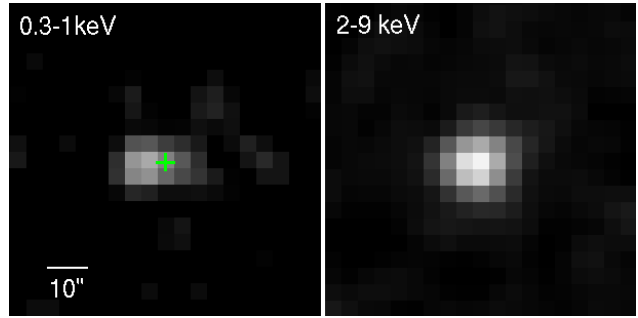}
\caption{
The EPIC pn and MOS combined images in the 0.3--1 keV and 2--9 keV
bands.North is up, East is to the left. The images have a pixel size of 4".
The green cross in the soft band image indicates the position
of the image centroid of the hard band source. The soft flux appears
somewhat offset from the hard, but statistically the positions are in agreement.}
\label{soft and hard images}
\end{figure}

The X-ray source is clearly rather hard, with many of the counts collected above 2 keV. Fig. 1 shows EPIC pn and MOS combined images of the source in the 0.3--1 keV and 2--9 keV bands. While the hard band image is centred at the optical position of the galaxy, the peak of the faint soft X-ray emission appears to be displaced to the east. by $\sim 5$", corresponding to $\sim 22$ kpc. The soft emission appears slightly extended, however, and the centroid of the emission is displaced by only 2.4". Fitting the PSF of the EPIC-pn to the soft and hard images we estimate the errors in the centroid positions to be 0.6" for the hard image and 2" for the soft, so the displacement is only marginally significant, and awaits verification by a better quality image. We note, however, that there might be a link to the optical superwind nebula traced by [OIII], which  extends $\sim 50$ kpc to the East and $\sim 10$ kpc to the West (Heckman, Armus \& Miley 1990); The R band image taken by Drake et al (2003) also shows an extension to the East. The soft X-ray emission is similar to these optical features in direction and size, and could therefore be associated with the superwind, rather than the HLIRG nucleus. 

\section{The EPIC spectrum}

\subsection{Hard band spectrum}

\begin{figure}
\includegraphics[width=50mm,angle=270]{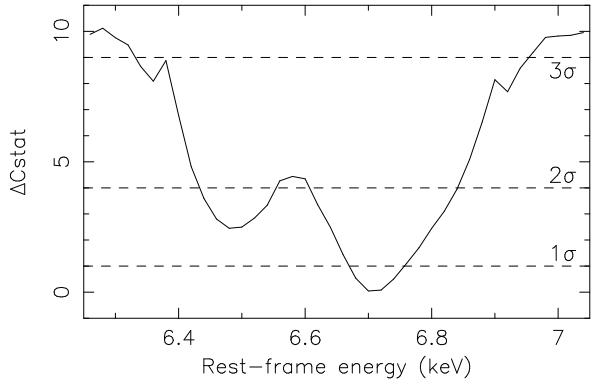}
\caption{Change in C-statistic versus rest frame line energy. The best fit indicates He-like iron emission at 6.7~keV, with neutral iron (6.4 keV) being strongly disfavoured. 
\label{fig:cstat}} 
\end{figure}

The X-ray spectrum obtained from the EPIC pn and MOS data is flat ($\Gamma\sim 0.2$) with a weak excess below 1 keV. We first consider the hard band $>2$~keV spectrum. Strong line emission at around 5 keV in the observers frame is seen, which can be identified with redshifted Fe K$\alpha$. 
Weak emission lines are sometimes seen in this energy range due to 
instrumental background. We have verified that the line feature is not a residual of the
instrumental Cr line at $\sim 5.4$ keV by inspecting three narrow band
images: 4.8--5.2 keV, which contains the redshifted Fe K$\alpha$ line feature, and two
neighbouring bands with the same energy width of 0.4 keV. The 4.8-5.2 keV image
shows a clear excess relative to the neighbouring bands. When modeled by
a single Gaussian, the centroid of the line is found at $6.70^{+0.08}_{-0.08}$ keV in the rest-frame (C-stat fit; errors here and henceforth are 90 per cent confidence for one parameter of interest, $\Delta C$ or $\Delta\chi^{2}=2.7$),  which indicates that the major component of the line is Fe {\sc xxv}, i.e. He-like iron. Cold Fe K$\alpha $ (less ionized than FeXVII) at 6.4 keV, which is usually found in
absorbed AGN, is ruled out at the $\sim 99$ per cent confidence level (Fig.~\ref{fig:cstat}).

The 2-10 keV continuum is hard and shows some curvature towards the soft X-ray  consistent with the effect of absorption. Formally,  the spectrum can be fitted with a hot thermal plasma, as long as strong absorption is allowed. Assuming solar metallicity (Anders \& Grevesse 1989), the MEKAL model gives a temperature of $7.1^{+3.7}_{-2.5}$ keV and \nH$=2.1^{+0.7}_{-0.5}\times 10^{23}$\psqcm,  as measured in the galaxy rest frame, with $\chi^2=8.3$ for 14 degrees of freedom. The Fe K$\alpha$ line is well fitted in this model. The absorption corrected bolometric luminosity of this thermal emission spectrum is $\simeq 2\times 10^{44}$\ergps. The luminosity already rivals luminous
AGN and exceeds the range of galaxy groups. In principle it could be produced
by a moderately rich cluster, but then we would expect the emission to be unabsorbed and
significantly extended, so this can be ruled out. High temperature thermal
plasmas like this are also present in the nuclei of starburst galaxies, 
e.g. NGC253 (Pietsch et al 2001, but see also Weaver et al 2002), 
and ULIRGs (e.g. Arp220; Iwasawa et al 2005). The luminosity in IRAS 00182--7112 is
much too large to be accounted for by these effects, with the above examples
showing luminosities of only $10^{39-41}$~erg s$^{-1}$. Even if we scale the
emission of Arp 220 by $L_{\rm IR}$ this predicts only $\sim10^{42}$\ergps
for IRAS 00182--7112, far below the observed
hard X-ray luminosity.

Given the large absorbing column, the source behind the
absorbing gas is likely to be located in a very compact region.
The implied column density agrees well with that of
the  obscuring material probed by the
mid-IR spectrum (see Discussion), so it is natural to locate the hard X-ray
source inside this region. The thickness of this warm
inner shell is $<0.03$ pc (Spoon et al. 2004), 
located at a distance of 0.5-2 pc (Lutz et al. 2004) so it appears to take the form of a thin shell. 
Conservatively, if we adopt a spherical geometry with a radius of 1 pc, the distance
appropriate for an obscuring torus (e.g., Krolik \& Begelman 1988) and consistent
with specific estimates for IRAS 00182--7112 itself
(Lutz et al. 2004), the gas density implied from the thermal spectrum is of the
order of $10^5$ cm$^{-3}$. This violates our assumption of an
optically thin plasma but in any case, such a dense plasma
would cool too fast ($\sim 200$ yr) to be observed. Given these
considerations, we consider it implausible that the emission is
thermal and is much more likely to be produced by an AGN. 

The hard continuum can be modeled as a power-law  with $\Gamma = 1.8$, a
typical value for Seyfert nuclei ignoring the effects of reflection,
and a cold absorbing column with
\nH $=2.0^{+0.8}_{-0.6}\times 10^{23}$\psqcm. 
The He-like Fe K line is unresolved when modelled with a narrow Gaussian,
and is very strong with a flux of $3.1^{+1.6}_{-1.6} \times 10^{-6}$ photon cm$^{-2}$ s$^{-1}$ corresponding to an equivalent width (EW) in the rest frame of $910\pm 460$~eV.  
A weaker line could be present at 6.4 keV but with its EW less
than 300 eV (90 per cent upper limit). 
The large EW of the He-like line indicates that
we are not seeing the direct emission from the central source, 
as such a large EW is only observed in a
reflection-dominated spectrum. The Fe K line centroid rules out cold
gas as the reflecting medium. This leads us to introduce reflection
from highly ionized matter. Here we use the model of Ross \&
Fabian (2005), which computes reflection spectra from the upper layer
of a slab of optically thick matter, like an accretion disk, illuminated by an
X-ray source. We assumed the illuminating source has a spectral
slope of $\Gamma = 2$ (appropriate for the intrinsic spectrum; Nandra \& Pounds 1994)
and the Fe abundance of the reflecting matter to be
solar. The 2--9 keV spectrum agrees well with the refection model
with an ionization parameter $\xi = L/(nR^2)\simeq 700$ erg cm
s$^{-1}$, modified by absorption of \nH $=1.8^{+0.8}_{-0.6}\times
10^{23}$\psqcm\ ($\chi^2 = 8.6$ for 14 degrees of freedom). The value
of $\xi$ is constrained mostly by the energy and strength of the Fe K feature and the 90 per cent
confidence range is $\xi = 320 - 3500$. The absorption-corrected rest
frame 2--10 keV luminosity is $0.8\times 10^{44}$\ergps.

\subsection{Soft band spectrum}

Because of the possible spatial displacement and its correspondence to the extended optical features, we think that the soft X-ray emission ($<2$ keV) is different in origin from the hard X-ray emission. By
analogy to the extended nebulae seen in nearby starburst galaxies, the soft X-rays are probably due to thermal emission with a temperature of $\sim 10^7$ K. A MEKAL thermal emission spectrum (Kaastra
1999) with $kT = 0.9^{+0.8}_{-0.2}$ keV and solar metallicity, modified only with Galactic absorption \nH $=4.2\times 10^{20}$\psqcm (Dickey \& Lockman 1990), gives a good fit to the 0.5-2 keV data, but the data quality means the parameters are very poorly constrained. An origin in extended photoionised gas is also possible, as seen in nearby Seyfert 2s (e.g. Guainazzi \& Bianchi 2007).  The rest-frame, absorption-corrected 0.5--2 keV luminosity is estimated to be $1.4\times 10^{42}$\ergps, which is slightly more luminous than classical nearby ULIRGs (Iwasawa 1999). It is about a factor 4 lower than the  $L_{\rm SX}/L_{\rm IR}$ ratio of nearby starburst galaxies (e.g., Ranalli, Comastri \& Setti 2003). 

\subsection{Broad band fit}

\begin{figure}
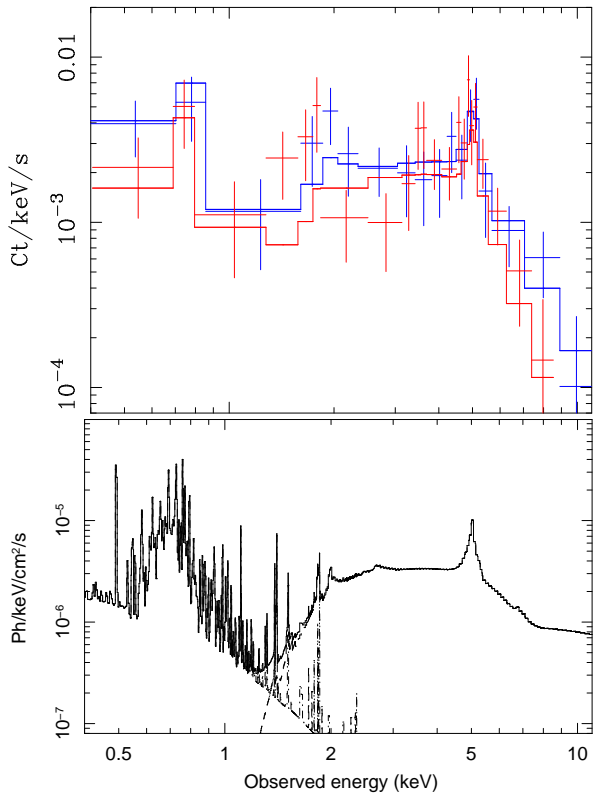
 
\hbox{\hspace{0.5mm}\includegraphics[width=54.5 mm,angle=270]{pnmosspec_col.ps}}
\includegraphics[width=50mm,angle=270]{mosp.ps}
\caption{
Upper panel: The full-band EPIC pn (blue) and MOS (red) spectra of IRAS 00182--7112
with the best-fit model as a histogram, which is folded through the
detector response. The model consists of optically--thin 
thermal emission for the soft X-ray emission and an ionized reflection
spectrum modified with internal absorption for the hard X-rays (see
text for details). The more lightly binned spectra (Section 2) are shown for display purposes. The model is plotted in the bottom panel and is an excellent
fit to the data.}
\label{fig:spec}
\end{figure}

The 1--2 keV region is the noisiest range of the spectrum and so far has been omitted from the spectral fits presented above. Considering this region we see possible evidence for a line-like excess at 2 keV. If thermal, this is is unlikely to be associated with the soft emission as the only plausible identification is with sulphur and would require a large overabundance ($>10$). An origin in photoionized gas seems more likely, as the feature could be a blend of SXVI Ly$\alpha$ and SiXIV radiative recombination continuum (RRC). In principle this could arise via the same mechanism as that  producing the Fe K$\alpha$ line, i.e. X-ray reflection, but with a lower $\xi$= 100-300 erg cm s$^{-1}$.  This seems reasonable because photoionized plasmas around AGN are expected to, and generally show a wide range of ionization parameter (e.g Krolik \& Kriss 2001;  Kinkhabwala et al 2002).  With this caveat, and to keep the spectral model simple, we have tested only a single ionized reflection component, together with the soft thermal emission, against the broad-band spectrum (Fig.~\ref{fig:spec}). The reflection spectrum is modified by absorption intrinsic to the object
and the thermal component is modified only by the Galactic extinction. The results hold for the soft thermal emission as in Section 4.2 but the ionization parameter of the reflection spectrum shifts to the intermediate value of $\xi =420^{+590}_{-120}$ erg cm s$^{-1}$, as expected, and the absorption column density decreases to \nH $\simeq 1.0^{+0.3}_{-0.3}\times 10^{23}$\psqcm. This two-component
model agrees with the broad-band data very well, with $\chi^{2}=21.4$ for 21 degrees of freedom.

\section{Discussion}

The XMM-Newton observation of IRAS 00182--7111 has revealed a very
luminous, hard X-ray source co-incident with the nucleus. The large X-ray 
luminosity observed in the hard band would normally be an
immediate indication of an AGN. However, given that the inferred
star formation rate in this object may be one of the highest in the universe,
further discussion is warranted to see whether stellar processes might account for the luminosity. 
In nearby star forming galaxies, the hard X-ray emission is generally attributed to X-ray binaries, 
and there is a roughly linear relationship between the hard X-ray luminosity and infrared
luminosity, and hence the star formation rate (e.g., David et al. 1992; Nandra et
al 2002; Ranalli et al 2003; Grimm et al 2003). Franceschini et al. (2003) have suggested that
young starburst systems  like ULIRGs  may follow a somewhat different relationship, but
the hard X-ray luminosity can still be used to estimate the star formation rate. 

Taking the most conservative of the published calibrations (Grimm et al. 2003), 
the star formation rate derived from the hard X-ray luminosity in IRAS
00182--7112 would be $>12,000 M_{\odot}$ yr$^{-1}$, 
while the star formation rate converted from the infrared luminosity (e.g.,
Kennicutt 1998) is about 1,000 $M_{\odot}$ yr$^{-1}$. This clearly
indicates that the hard X-rays are overluminous for a starburst
alone. Moreover, the intense, narrow, He-like Fe K line rules out X-ray
binaries as the major component of the hard X-ray emission.  Thermal emission 
from hot gas is highly unlikely, as discussed
in Section 4.1, so the evidence overwhelmingly points to an AGN
origin for the hard X-ray emission. 

The X-ray spectrum of  IRAS 00182--7111 is characterised by a flat, hard continuum
and intense iron emission. These are both tell-tale signs of a hidden, Compton--thick
AGN (Reynolds et al. 1996; Matt et al. 2000). Remarkably, we find in this case that the 
reflection-dominated spectrum arises from very highly ionized gas, with
iron predominantly in the He-like Fe {\sc xxv} state. Fe {\sc xxv} and even 
Fe {\sc xxvi} have been seen in Seyfert 2s (e.g., NGC1068, Marshall et al 1993; 
Ueno et al 1994; Iwasawa, Fabian \& Matt 1997), but the dominant feature in
Compton thick  AGN is always the cold 6.4 keV line.  As far as we are aware,
IRAS 00182--7111 is unique among obscured AGN in exhibiting a reflection--dominated
spectrum from ionized gas, so the origin of the line emission in this object is probably quite different to that in Seyfert 2s. 

Having established the presence of an AGN, it is natural to assess
its contribution to the overall energy budget. The
directly-viewed, absorption-corrected 2-10 keV luminosity is $L_{2-10}^{\rm refl}\simeq
0.8\times 10^{44}$\ergps, but this is only the reflected light of a
hidden illuminating source. The reflection
efficiency depends on the ionisation parameter, 
and for a 2-10 keV albedo $\epsilon $, the illuminating
luminosity required to produce the observed reflection is $L_{2-10}^{\prime} =
\epsilon^{-1}L_{2-10}^{\rm refl}$
The appropriate albedo for an ionized reflecting medium 
can be estimated from the work of Ross \& Fabian (2005), yielding $L_{2-10}^{\prime} \sim 2\times 10^{44}$\ergps. It is unlikely that all the
illuminated, ionized medium is visible to us. This introduces a geometrical
correction factor of $(\Omega/2\pi)^{-1}$ where $\Omega $ is the solid
angle that the ionized reflecting medium occupies in the sky viewed
from the source but also visible to the observer. While we have assumed
an optically thick medium for the calculation of the reflection spectrum, the 
optical depth of the reflecting medium is in reality unknown, and 
if $\tau_{\rm T}<1$ the reflection is reduced correspondingly. If
these corrections are included in a factor $\eta $, the illuminating
source has $L_{2-10}=\eta^{-1}L^{\prime}_{2-10}$. The bolometric
correction, $f_{\rm HX}=L_{2-10}/L_{\rm bol}$, is typically 0.05
for high-luminosity AGN (e.g., Elvis et al. 1994). The
bolometric AGN luminosity is then estimated as $L_{\rm bol}^{\rm
AGN}=f_{\rm HX}^{-1}\eta^{-1}L^{\prime}_{2-10}\sim 2\times
10^{46}(f_{\rm HX}/0.05)^{-1}(\eta/0.2)^{-1}$ erg s$^{-1}$, or 
$5 \times 10^{12}$~$L_{\odot}$
for the fiducial values of $f_{\rm HX}$ and $\eta$. 
Our best estimate is therefore that the AGN
dominates the total FIR luminosity of $\sim 7\times 10^{12}L_{\odot}$, albeit with very large uncertainties. 

Other evidence supports this conclusion. The properties of the soft X-ray  emission below 1 keV are consistent with thermal emission powered by a starburst, though a contribution from photoionized gas cannot be excluded with the present data quality. It is
plausibly offset and/or extended from the nucleus in a manner 
co-incident with the extended optical emission, and the spectrum is soft
and unabsorbed. The inferred star-formation
rate from the soft X--ray luminosity (Ranalli et al. 2003) is 
$310 M_{\odot}$ yr$^{-1}$. This is
lower than the FIR-inferred star formation rate, and therefore consistent with
the idea that there is another major contributor to the FIR power.  
This is also indicated by the mid-IR data. 
We have already mentioned that the Spitzer spectrum is dominated by absorption 
and shows weak PAH emission. The 
star formation linked to the PAH emission must be located outside the
central obscuration, and that star-formation rate derived from the PAH luminosity
accounts for only up to 30 per cent of
the bolometric luminosity (Spoon et al 2004). This is in excellent agreement
with the soft X-ray derived star formation rate.

On the face of it, the ionized emission line seen in IRAS 00182--7111 is at odds with the observed flat continuum shape. With Fe in the He-like state, the lighter elements able to suppress the soft X-rays in the reflecting medium will be mostly stripped.  In our spectral modeling, we account for this by an additional absorption column of \nH $\sim 10^{23}$\psqcm\ in the line of sight. This can be identified with the same matter absorbing the mid-IR continuum. Spoon et al (2004) identified two layers of obscuring material: a
cold, outer shell responsible for the ice and silicate absorption and a warm, inner (gas-)shell inferred from the CO absorption band at 4.67$\mu $m. The properties of these shells of particular interest to us are the thickness of the inner shell, which is constrained to be smaller than 0.03 pc, its location at $\sim 0.5-2$~pc and the estimated equivalent hydrogen column density of the order of $10^{23}$\psqcm\ for both shells. The small size of the inner shell, and the the total column density agrees perfectly with that needed to provide the necessary suppression of the nuclear (reflected) soft X-rays. If we are correct in our assessment that IRAS 00182-7111 hosts a Compton thick AGN, it implies that the line--of--sight to the nucleus is much more heavily obscured than those to both the source of the reflected X-rays, and the mid-IR emission. It is also worth noting that the fact that we can see the hot dust
means that not all the directions from the central source are obscured
as heavily as the line of sight. This argues for a torus-type geometry for
the IR absorber. A combination of detailed X-ray and mid-IR observations can therefore help disentangle the geometry of the very innermost regions.

In Compton thick Seyfert 2s, the obscuration and reflection is associated 
with the torus envisaged in orientation-dependent unification
schemes (e.g. Antonucci \& Miller 1985). This 
is an obvious identification in IRAS  00182--7111, and 
it is tempting to associate the high ionization seen in IRAS  00182--7111 with its
hyper-luminous classification. Intense illumination from the powerful central source
might ionize the surface of any obscuring torus. In this regard it is worth considering
the constraints on the physical condition of the reflecting medium. Taking an
ionizing luminosity $L_{ion}\simeq (1/2)L_{\rm bol}^{\rm AGN} = 1\times
10^{46}$\ergps, a density $n=10^6$ cm$^{-3}$, similar
to that of the mid-IR CO absorption layer (Spoon et al 2004),
and ionization parameter $\xi = L_{ion}/(nR^2)\sim 600$ erg cm
s$^{-1}$, the distance of the reflecting medium from the central
source can be expressed as $R\sim 1 (n/10^6{\rm
cm}^{-3})^{-1/2}(\xi/600{\rm erg\thinspace s\thinspace
cm}^{-1})^{-1/2}(L_{ion}/1\times 10^{46}{\rm erg s}^{-1})^{1/2}$
pc. This is consistent with a torus origin (Lutz et al. 2004).  If the depth of the medium 
is assumed to be $l\simeq R$, the
reflection column density is $nl\sim 4\times 10^{24}(n/10^6{\rm
cm}^{-3})$cm$^{-2}$, or roughly $2\tau_{\rm T}$. This is also
a plausible optical depth for the torus. 
%The optically thick condition
%we assumed for the calculation of the reflection spectrum also seems reasonable.

An alternative is to associate the reflector with hot
gas filling the torus, and in this case it may have a lower optical depth. Krolik
\& Kallman (1987) have predicted ionized iron line emission from this medium, and
it is seen in local Seyfert 2 galaxies such as NGC 1068 (Marshall et al. 1993; Iwasawa et al. 1997).
It is difficult to see how such a medium would be confined, and one possibility is that we
are seeing X-ray reflection from a hot outflow. The superwind seen in the optical may in fact be an AGN-driven outflow, and the soft X-ray emission could arise either via radiative heating, or shock heating of the interstellar medium by this wind.  Such AGN-driven winds have received much recent attention as they can provide the necessary link between black hole accretion and bulge formation via AGN feedback (e.g. Silk \& Rees 1988). Several authors have suggested evolutionary links between ULIRGs and AGN (e.g. Sanders et al. 1988). For example, recent numerical simulations suggest that massive galaxies undergo violent merger-driven star formation and obscured AGN activity (Hopkins et al. 2005). Eventually, accretion powered winds can sweep up gas from the galaxy, terminating star formation and leaving a red bulge-dominated remnant. IRAS  00182--7112 may represent an example of a galaxy undergoing its obscured accretion phase, and be in the process of quenching its star formation. Future high resolution X-ray spectroscopy (with, e.g. XEUS or Constellation-X) could reveal shifts and profile distortions in both the iron K$\alpha$ line and the soft X-ray emission lines characteristic of such an outflow. 
 
The X-ray detection of an AGN in IRAS 00182--7112 could have important implications for surveys of HLIRGs at high redshift. Houck et al (2005), for instance, found a number of HLIRG candidates at $z>2$ in a Spitzer survey of sources with MIPS 24 $\mu$m detection. They argued that many of them were powered predominantly by AGN, largely relying on the similarity of their IRS  mid-IR spectra to that of IRAS 00182--7112. The fact that we now have robust evidence for a powerful AGN in IRAS 00182--7112 argues that these other HLIRGs do as well. 

Based on optical emission line diagnostics, as many as 50~per cent of HLIRGs are thought to host an AGN. The optical classification of IRAS 00182--7112 is that of a LINER (Almus et al. 1987), hinting that there may be a large additional population of AGN-hosting HLIRGs without obvious optical, or for that matter mid-IR, AGN signatures. In fact, sufficiently deep X-ray observations could in principle reveal a buried AGN in all HLIRGs. 

The connection between AGN-dominated HLIRGs and submillimetre
galaxies (SMGs) at high-redshift is unclear. While AGN do seem to be common
in SMGs, the AGN contribution to the total energy budget appears 
to be minor (Alexander et al 2005a). In this respect, one archetypal HLIRG, the gravitationally-lensed IRAS F10214+4724 at $z=2.3$ (Rowan-Robinson 2000) appears more like a SMG in nature (Alexander et al 2005b). It should be noted, however, that our estimate of the bolometric contribution of 
the AGN in  IRAS 00182--7112 relies on a relatively high quality X-ray spectral
decomposition, which is not yet possible for most  SMGs. Very sensitive future X-ray observation are required for a definitive test of whether these too are mainly powered by AGN, or starbursts.  

%HLIRGs are rare but as the number of HLIRGs discovered by Spitzer
%increases, it becomes clear that the environment is an important
%factor for the formation of HLIRGs, since they are not isolated but
%usually found within cosmic large scale structures at high redshift
%(Ref). The IRAS-dicovered HLIRGs, IRAS 09104+4109 ($z=0.44$) and IRAS
%F15307+3252 ($z=0.93$) are both found to be the central galaxies of a
%rich cluster (Kleinmann et al 1988) or a group/poor cluster (Iwasawa
%et al 2005). For IRAS 00182--7112 however, no evidence that the galaxy
%lives in an over-density region has been found so far.

\section*{Acknowledgements}
We gratefully acknowledge the efforts of those we built and operate the {\it XMM-Newton} satellite. We thank the anonymous referee, Henrik Spoon and Michael Rowan-Robinson for helpful comments on the manuscript. KN thanks the Leverhulme Trust for financial support.

\label{lastpage}

\end{document}